\pdfoutput=1

\documentclass[11pt,twoside,a4paper,cmspaper,final,collab]{cms-tdr}
\def\svnVersion{175797}\def\cmsCernNo{2012-309}\def\cmsCernDate{\today}\def\cmsMessage{Submitted to the Journal of High Energy Physics}

\begin{document}\cmsNoteHeader{B2G-12-004}

\hyphenation{had-ron-i-za-tion}
\hyphenation{cal-or-i-me-ter}
\hyphenation{de-vices}

\RCS$Revision: 166586 $
\RCS$HeadURL: svn+ssh://svn.cern.ch/reps/tdr2/papers/B2G-12-004/trunk/B2G-12-004.tex $
\RCS$Id: B2G-12-004.tex 166586 2013-01-23 18:16:18Z skhalil $
\newlength\cmsFigWidth
\ifthenelse{\boolean{cms@external}}{\setlength\cmsFigWidth{0.85\columnwidth}}{\setlength\cmsFigWidth{0.4\textwidth}}
\ifthenelse{\boolean{cms@external}}{\providecommand{\cmsLeft}{top}}{\providecommand{\cmsLeft}{left}}
\ifthenelse{\boolean{cms@external}}{\providecommand{\cmsRight}{bottom}}{\providecommand{\cmsRight}{right}}
\cmsNoteHeader{B2G-12-004} 
\title{Search for heavy quarks decaying into a top quark and a W or Z boson using lepton + jets events in \Pp\Pp\ collisions at $\sqrt{s} = 7$\TeV}

\date{\today}

\abstract{
Results are presented from a search for the pair-production of heavy
quarks, $\mathrm{Q\overline{Q}}$, that decay exclusively into a top quark
and a W or Z boson. The search is performed using a sample of proton-proton
collisions at $\sqrt{s} = 7$\TeV corresponding to an integrated
luminosity of 5.0\fbinv, collected by the Compact Muon Solenoid
experiment. The signal region is defined using a sample of events containing
one electron or muon, missing transverse momentum, and at least four jets with
large transverse momenta, where one jet is likely to
originate from the decay of a bottom quark. No significant excess of events
is observed with respect to the standard model expectations. Assuming
a strong pair-production mechanism, quark masses
below 675 (625)\GeV decaying into tW (tZ) are excluded at the 95\% confidence level.
}

\hypersetup{%
pdfauthor={CMS Collaboration},%
pdftitle={Search for heavy quarks decaying into a top quark and a W or Z boson using lepton + jets events in pp collisions at sqrt(s) = 7 TeV},%
pdfsubject={CMS},%
pdfkeywords={CMS, physics, heavy quarks}}

\maketitle 

\section{Introduction}
\label{sec:introduction}

New heavy quarks ($\mathrm{Q}$) that decay to top quarks and electroweak bosons ($\mathrm{W}$, $\mathrm{Z}$,
or Higgs bosons) appear in many theoretical scenarios, the simplest being the sequential
fourth generation model~\cite{4Gen}. Experimental constraints on masses of the
fourth-generation quarks~\cite{D0,CDF,DirectMass} suggest that the dominant decay mode of the down-type
fourth-generation quark is to a top quark and a $\mathrm{W}$ boson.
There are also models~\cite{compositeHiggs,ExtraDimensions,phenomLittleHiggs}
that include non-chiral heavy quarks with vector-like couplings to bosons.
Such quarks cancel quadratically divergent corrections to the Higgs boson mass, and thereby
stabilize it 
at the electroweak symmetry breaking scale. In these models, the new quarks
can also decay through flavor-changing neutral-current processes, such as to a top quark and a
$\mathrm{Z}$ boson, or to a top quark and a Higgs boson.

This Letter presents a search for the pair production of
new heavy quarks $\mathrm{Q}$
that decay exclusively to
a top quark and a $\mathrm{W}$ boson,
or
to a top quark and a $\mathrm{Z}$ boson.
The current experimental constraints on masses of heavy quarks $\mathrm{Q}$
are set by previous LHC searches. If $\mathrm{Q}$ exists, its mass must be greater
than 480\GeV, assuming the exclusive decay $\mathrm{Q\to tW}$~\cite{ATLAS_bprime}, or
greater than 475\GeV, assuming the decay $\mathrm{Q\to tZ}$~\cite{CMS-tZ}.
Given these constraints, the mass splitting between the heavy quark $\mathrm{Q}$ and the top quark
must be large, so that the decay products can be produced on-shell.
For heavy down-type quarks decaying exclusively into a top quark and a $\mathrm{W}$ boson,
the full decay chain
is
$\mathrm{Q\overline{Q}} \to \mathrm{tW}^-\mathrm{\overline{t}W}^+ \to \mathrm{bW}^+\mathrm{W}^-\mathrm{\overline{b}W}^-\mathrm{W}^+$, and for up-type quarks decaying exclusively into a $\mathrm{t}$ quark and a $\mathrm{Z}$ boson,
it is
$\mathrm{Q\overline{Q}} \to \mathrm{tZ}\mathrm{\overline{t}Z} \to \mathrm{bW}^+\mathrm{Z}\mathrm{\overline{b}W}^-\mathrm{Z}$.
The search is performed in events in which one of the $\mathrm{W}$ bosons
(originating either from the decay of the heavy quark, or from the subsequent decay of a top quark)
decays leptonically, while the
other bosons decay into quark-antiquark pairs. Selected events are required to have exactly one charged lepton,
an imbalance in transverse momentum (\PTm), and at least
four jets with high transverse momenta (\pt), at least one of which is consistent with the decay of
a bottom quark.

The dominant standard-model (SM) processes that result in the same signature include
$\ttbar$ production, as well as production of $\mathrm{W}$ bosons with associated jets.
These background processes are characterized by smaller lepton and jet transverse momenta
and lower jet multiplicities than those in heavy quark decays.
The search for heavy quarks is performed by classifying events
based on the number of final-state jets. For each jet multiplicity,
the scalar sum ($S_{\mathrm T}$) of the transverse momenta of the lepton, the jets, and
\PTm are used
to test for the presence of a new physics signal in the data.
Signatures with high jet multiplicity and large values of $S_{\mathrm T}$ are predicted in a variety of new physics scenarios~\cite{strassler},
making the search presented in this Letter sensitive to a broad class of models of new physics.

\section{The CMS Detector and Data Samples}
\label{sec:data}

The data were recorded during 2011 by the Compact Muon Solenoid (CMS)
experiment at the LHC,
which delivered proton-proton collisions at $\sqrt{s}=7$\TeV, and correspond
to an integrated luminosity of 5.0\fbinv.
The CMS detector uses a polar coordinate system with the $z$ axis pointing along
the counterclockwise circulating beam. The $x$ axis points towards the center
of the Large Hadron Collider (LHC) ring, and the $y$ axis points up.
Angular coordinates are specified by the pseudorapidity $\eta = -\ln[\tan(\theta/{2})] $,
where $\theta$ is the polar angle measured with respect to the positive $z$ axis, and by $\phi$,
the azimuthal angle about this axis.

A characteristic feature of the CMS detector is its superconducting
solenoid, which is 6\unit{m} in diameter and 13\unit{m} in length, and provides an axial
magnetic field of 3.8\unit{T}. Located inside the solenoid is a
multilayer silicon-pixel and silicon-strip tracker, covering the
pseudorapidity region $|\eta|<2.5$, the electromagnetic calorimeter (ECAL), covering
$|\eta|<3.0$ made of lead-tungstate crystals, the preshower detector covering
$1.65<|\eta|<2.6$, and the hadronic calorimeter (HCAL) made of brass and
scintillators, covering $|\eta|<3.0$. Muons are measured with gas-ionization detectors
embedded in the return yoke of the solenoid, which cover $|\eta|<2.4$,
and complement the measurement in the inner tracking detector.
The CMS detector is nearly hermetic,
allowing for momentum balance measurements
in the plane transverse to the beam direction.
A two-tier trigger
system selects the most interesting pp collision events for use in physics analysis.
A detailed
description of the CMS detector is given in Ref.~\cite{CMS}.

The $\mathrm{e}$+jets events were collected with triggers requiring at least
one electron candidate, with a \pt threshold ranging from 25 to 32\GeV.
When the LHC instantaneous luminosity increased,
at least three jets with $\pt>30$\GeV were also required.
The $\mu$+jets events were
collected with triggers that required at least one muon candidate with
a \pt threshold ranging between 30 and 40\GeV.
The presence of jets was not required in
the triggers for the $\mu$+jets events, and no conditions were placed on
the \PTm for either channel.

The background processes $\ttbar$+jets, $\mathrm{W}$+jets, and $\mathrm{Z}$+jets, are simulated
using the {\MADGRAPH~5.1.1} event generator~\cite{madgraph} with {CTEQ6L1}~\cite{cteq6m} parton distribution
functions (PDF). The single top quark production via $\mathrm{tW}$, $s$ and $t$ channels is simulated
using the {\POWHEG~1.0} event generator~\cite{powheg1,powheg2,powheg3} with {CTEQ6M} PDF~\cite{cteq6m}.
The multijet and diboson processes ($\mathrm{WW}$, $\mathrm{WZ}$, and $\mathrm{ZZ}$) are
generated using the {\PYTHIA~6.424} event generator~\cite{pythia} with {CTEQ6M} PDF.
{\PYTHIA} is also used to model the parton shower and hadronization
for both {\MADGRAPH} and the {\POWHEG} Monte Carlo
(MC) samples.
The generated events are processed through
a CMS detector simulation based on {\GEANTfour}~\cite{geant4}.
Additional minimum-bias events (pileup) are
generated with {\PYTHIA} and superimposed on the hard-scattering events to simulate multiple
collisions within the same bunch crossing. All the MC simulated events are weighted to reproduce the distribution of the
number of interaction vertices observed in data.

\section{Event Reconstruction}
\label{sec:eventReconstruction}

Events are reconstructed using the CMS particle-flow (PF) algorithm~\cite{particleflow,particleflow1,particleflow2},
which identifies all observable particles in an event by combining the information from
charged particles in the silicon tracker, energy deposited in the ECAL and HCAL, and
signals in the preshower detector and the muon systems. This procedure separates all particles
into five categories: muons, electrons, photons, and charged and neutral hadrons. Energy
calibration is performed separately for each particle type. The imbalance in transverse 
momentum \PTm in an event is defined as the negative vector sum
of the transverse momenta of all objects from the PF algorithm. Events must also have an acceptable
primary vertex, and we select the vertex with the largest value for the scalar sum
of the $\pt^2$ of the associated tracks.

Electron candidates are reconstructed from clusters of energy deposited
in the ECAL. The clusters are first matched to
track seeds in the pixel detector. The track trajectories of electron candidates are reconstructed
using a 
dedicated modeling of the electron energy loss, and fitted with a Gaussian-sum
filter~\cite{CMS_NOTE_2005-001}.

Muon candidates are identified through different reconstruction algorithms
using hits in the central silicon tracker and signals in the outer muon
system~\cite{MuonRecons}. A standalone muon algorithm uses only
information from the muon chambers. The tracker muon algorithm begins with
tracks found in the inner tracker, and associates these with matching segments
in the muon chambers. In this analysis, all muons have to pass the global muon
algorithm, which starts off with the standalone muons and then performs a
global fit to the hits in the tracker and the muon system for each muon candidate.

Jets are reconstructed using the anti-\kt jet
clustering algorithm~\cite{antikt} with a distance parameter $R=0.5$, as
implemented in \textsc{Fastjet} version {2.4}~\cite{Cacciari:2005hq,fastjet,fastjet1,fastjet2}. Jets are
identified as originating from the decay of a bottom quark through
the combined secondary vertex (CSV) algorithm at the medium operating point~\cite{CMS-PAS-BTV-11-003}.
The CSV algorithm provides optimal $\mathrm{b}$-tagging performance by combining information
on the impact parameter significance, the properties of the secondary vertex, and the jet kinematics.
The variables are combined using a likelihood-ratio technique to compute a
$\mathrm{b}$-tagging discriminant. The residual differences in the performance of
the $\mathrm{b}$-tagging algorithm between data and simulation are accounted for by
\pt- and $\eta$-dependent data/simulation scale factors~\cite{CMS-PAS-BTV-11-003}.

\section{Event Selection}
\label{sec:selection}

Charged leptons from the decay of $\mathrm{W}$ bosons are typically well isolated from jets.
The lepton isolation can be expressed in terms of the quantity $I^\ell_r$,
defined as the scalar sum of the $\pt$ 
of charged hadrons, neutral hadrons and photons in a cone of
$\Delta R=\sqrt{(\Delta \phi)^2+(\Delta \eta)^2}<0.3$ around the lepton momentum vector, divided by the
lepton \pt. The isolation requirements are optimized
to be $I^\Pe_r<0.1$ for electrons, and $I^{\mu}_r<0.125$ for muons. The electrons (muons)
also must have $p^\Pe_{\mathrm T}>35$\GeV ($p^{\mu}_{\mathrm T}>42$\GeV), and $|\eta^{\Pe}|<2.5$
($|\eta^{\mu}|<2.1$). The lepton trajectories are required to have a magnitude of the transverse impact
parameter less than 0.02\unit{cm} and a magnitude of the longitudinal impact parameter along the beam direction less than 1\unit{cm} relative to the primary vertex.

The final selection requires events to have exactly one isolated lepton and at least four
jets with $|\eta|<2.4$ and $\pt>$ 100, 60, 50, 35\GeV. 
Additional jets having $\pt >$ 35\GeV are also counted.
The minimum number of jets, and the jet \pt requirements
are optimized to enhance the sensitivity to the $\mathrm{Q\overline{Q}}$ signal.
The thresholds for lepton
$p^l_{\mathrm T}$ and the third jet \pt are driven by trigger conditions.
Jets that are within a cone of $\Delta R<0.3$ of the lepton direction
are ignored. At least one jet must be b-tagged by the CSV algorithm.
The event is also required to have $\PTm > 20$\GeV.

Table~\ref{tab:events} lists the number of events observed and the number expected for
the background sources, following all selections. The cross section for $\ttbar$
production is taken from Ref.~\cite{ttbar_xsec}.
The single top quark cross
sections are approximate NNLO calculations obtained from Ref.~\cite{ST1,ST2,ST3}.
The cross sections for $\mathrm{W}$+jets and
$\mathrm{Z}$+jets are computed to NNLO using \textsc{fewz}~\cite{FEWZ}.
The cross sections for the diboson processes $\mathrm{WW}$, $\mathrm{WZ}$, and $\mathrm{ZZ}$
are calculated to NLO using \textsc{mcfm}~\cite{MCFM}.
The expected number of SM background events is
evaluated based on the cross sections given in Table~\ref{tab:events},
the corresponding efficiencies and acceptances for each background, and integrated luminosity,
with exception of the contributions from multijet processes, which are estimated from data.
We perform a fit of SM contributions, normalized as described above, to the
$\PTm$ distribution in data. In the fit, we constrain the $\ttbar$+jets
contribution within its measured uncertainty, while Z+jets, single top and diboson
production processes are constrained within their theoretical uncertainties.
The normalizations for the W+jets and multijet contributions are allowed to float freely.
We obtain multijet scale factors as a function of the lepton $\eta$ and use them to correct
for the multijet normalization from simulation. For $\mu$+jets, the multijet background using this technique
is found to be negligible.

Table~\ref{tab:signal} presents the $\mathrm{Q\overline{Q}}$ production cross sections
that are computed at approximate NNLO using \textsc{hathor}~\cite{HATHOR},
along with the expected number of events for the
$\mathrm{e}$+jets and $\mu$+jets channels.

\begin{table}[htbp]
\topcaption{Background cross sections, expected number of background events and observed number
of events in 5.0\fbinv data, for the \Pe+jets and $\mu$+jets samples prior to the
likelihood fit. The uncertainties on the expected number of events reflect only the statistics of the simulation.
The quoted uncertainties on the cross sections are theoretical or measured in the case of $\ttbar$+jets.
}
\label{tab:events}
\begin{center}
\begin{tabular}{lccc}
\hline\hline
Background process &    Cross section   &          $\mathrm{e}$+jets events                    &       $\mu$+jets events        \\
\hline
$\mathrm{t\overline{t}}$+jets &  $154\pm19$ pb         & ~$7521\pm38$~                &  ~$7190\pm37$ \\
Single top &  ~$84.9\pm2.5$ pb           & $399\pm4$                    &  ~~$391\pm4$~~ \\
W+jets          &  ~$31\pm2$ nb           & $798\pm16$                   &  ~$790\pm16$~ \\
Z+jets          & ~~~$3.1\pm0.3$ nb          & ~$104\pm3$                   &  ~$63\pm2$ \\
Diboson (WW, WZ, ZZ)  &  ~$67\pm2$ pb          & ~$17\pm1$                   &  ~$15\pm1$ \\
Multijet        &          from data             & ~$334\pm8$                  &  --       \\
\hline
Total background &                      & ~$9173\pm42$                 &  $8449\pm41$ \\
\hline
Data            &                       & 9109                         & 8211 \\
\hline\hline
\end{tabular}
\end{center}
\end{table}

\begin{table}[htbp]
\topcaption{Signal cross sections~\cite{HATHOR}  
and expected number of $\mathrm{Q\overline{Q}}$ signal events in the $e$ and $\mu$ channels for four
quark masses. The uncertainties reflect the statistics of the MC simulations.}
\label{tab:signal}
\begin{center}
\begin{tabular}{lccccc}
\hline\hline
$M_{\mathrm Q}$ (GeV) & $\sigma$ (pb)&  $ \mathrm{Q \to tW}$   & $ \mathrm{Q \to tW}$ & $ \mathrm{Q \to tZ}$   & $ \mathrm{Q \to tZ}$  \\
  & & $\mathrm{e}$+jets & $\mu$+jets & $\mathrm{e}$+jets & $\mu$+jets \\
\hline
500 & 0.33   & $136.9\pm1.4$ & $137.0\pm1.4$ & $91.2\pm1.2$ & $85.9\pm1.1$ \\
550 & 0.17   & $74.6\pm0.8$ & $74.9\pm0.8$   & $49.1\pm0.6$ & $46.7\pm0.6$ \\
600 & 0.092  & $41.2\pm0.4$ & $41.7\pm0.4$   & $27.1\pm0.3$ & $26.1\pm0.3$ \\
650 & 0.051  & $22.9\pm0.2$ & $23.4\pm0.2$   & $15.9\pm0.3$ & $15.1\pm0.3$ \\
\hline\hline
\end{tabular}
\end{center}
\end{table}

\section{Likelihood Fit and Systematic Uncertainties}
\label{sec:limit}

The search for a $\mathrm{Q\overline{Q}}$ signal is performed
by fitting the data to the distribution of $S_{\mathrm T}$ as a function of jet multiplicity
($N_J$). The fit is performed for the combination of $\mathrm{e}$+jets and $\mu$+jets
channels and for $N_J$ =  4, 5, 6, and $\geq 7$ jets.
The bins are chosen so that the MC statistical uncertainty in
each bin is less than $17$\%. 

The dominant SM background is from $\ttbar$ production. Because the jet multiplicity
for the $\ttbar$+jets events is not well modeled, the  $\ttbar$ contributions for
$N_J$ = 4, 5, 6, and $\geq 7$ are allowed to
vary independently in the fit. A log-normal constraint is imposed on the expected yield
of $\ttbar$ events for each jet multiplicity sub-sample.
The normalization uncertainty for each sub-sample is determined from the difference
that results from changing the renormalization and factorization scales
by a factor of two relative to the nominal value
equal to $Q = \sqrt{m_{\mathrm t}^2 + \sum p_\mathrm{ T, jet}^2}$ , where the sum is taken over jets
produced in association with the $\ttbar$ pair. The inclusive
top quark pair production cross section $\sigma_{\ttbar}$ and its uncertainty are
taken from the recent CMS measurement~\cite{ttbar_xsec}. 
The uncertainty on $\sigma_{\ttbar}$ is used in the fit as a log-normal constraint
correlated between different jet multiplicities.

Other SM contributions include electroweak processes:
$\mathrm{W}$+jets, $\mathrm{Z}$+jets,
single top quark, and diboson production, as well as multijet events.
They are combined into a single background template.
The sum of these backgrounds is allowed to vary independently
across each jet multiplicity sub-sample, with
an uncertainty of 50\% assigned to the normalization of each sub-sample.

The luminosity is constrained to a log-normal distribution with an uncertainty of
2.2\%~\cite{Lumi}. The electron and muon trigger and identification efficiencies are obtained from data using
dilepton decays of $\mathrm{Z}$ bosons.
A conservative systematic uncertainty of 3.5\% is attributed to account for pileup and lepton $\eta$ dependence.
These efficiencies together with their uncertainties are treated as normalization constraints,
and applied to the electron and muon events respectively.

In addition to constraints on normalization, there are other parameters that affect
both the normalization and the shape of $S_{\mathrm T}$ and the jet multiplicity spectra. These include
the jet energy scale, the $\cPqb$-tagging efficiency, the matching between matrix element
partons and parton showers and the renormalization and factorization scales.
We incorporate these uncertainties in the fit by generating
additional templates corresponding to shifts by $\pm$ 1 standard deviation on the parameter in
question. The energies of the jets are corrected using the calibration constants
determined in Ref.~\cite{jet_calibration} as a function of \pt and $\eta$.
The uncertainties on $\mathrm{b}$-tagging
efficiencies are estimated by changing the $\mathrm{b}$-tagging efficiency by $\pm$1
of its standard deviation~\cite{CMS-PAS-BTV-11-003}.
The uncertainty due to the choice of factorization and renormalization scales is estimated by simulating two sets
of $\ttbar$ samples in which both scales are increased or decreased by a factor of two relative
to their nominal value.
The uncertainty arising from matching matrix element partons with parton
showers is estimated using two $\ttbar$ simulated samples, with matching threshold
shifted up or down by a factor of two relative to its default value (40\GeV). Other sources
of systematic uncertainties, such as jet energy resolution, \PTm resolution
and pileup interactions have negligible impact on the limit for a $\mathrm{Q\overline{Q}}$ signal.

Systematic uncertainties enter the likelihood through ``nuisance'' parameters
~\cite{asymptoticCLs}, that reflect the presence of imprecisely determined quantities
 that affect the $S_{\mathrm T}$ and jet multiplicity distributions. These are represented
by resolution functions contained within the likelihood function, and are integrated
over in the process of minimization,
resulting in a reduction in the final systematic uncertainty.

Table~\ref{tab:sys_summary} summarizes the systematic uncertainties included in the fit
to $S_{\mathrm T}$ and $N_J$. Parameters labeled ``Distribution'' affect both shape and normalization
of the $S_{\mathrm T}$ and $N_J$ distributions, where only their effect on background normalization is quoted.
Parameters labeled ``Normalization'' affect only the normalization of SM backgrounds and/or new physics
signal.

\begin{table}
\topcaption[Summary Systematics Table]{\label{tab:sys_summary}
List of systematic uncertainties included in the likelihood fit.
Parameters labeled ``Distribution'' affect both shape and normalization
of the $S_{\mathrm T}$ and $N_J$ distributions. The quoted uncertainties correspond
to their effect on normalization only.
}
\begin{center}
\begin{tabular}{ccc}
\hline\hline
 Parameter type & Source &   Uncertainty~(\%)  \\
\hline
\multirow{4}{*}{Distribution} & $\mathrm{Q^2}$ scales for $\ttbar$+jet & $ 8.7 $  \\
& Matching partons                         &     $ 5.8 $ \\
& Jet energy scale                         &     $ 5.4 $ \\
& b-tagging efficiency                         &     $ 5.1 $ \\
\hline
 \multirow{7}{*}{Normalization}  & Lepton ID/reco/trigger    & $ 3.5 $ \\
 & Luminosity                                    &  $2.2$ \\
& $\ttbar$ cross section                         &  $12$ \\
& $\mathrm{N_{jets}} = 4$   &  $3.5$ \\
& $\mathrm{N_{jets}} = 5$   &  $16$ \\
& $\mathrm{N_{jets}} = 6$   &  $23$ \\
& $\mathrm{N_{jets}} \geq 7$ &  $22$ \\
& Other backgrounds &  $50$ \\
\hline\hline
\end{tabular}
\end{center}
\end{table}

\begin{figure}[htp]
\begin{center}
\includegraphics[width=1.0\textwidth]{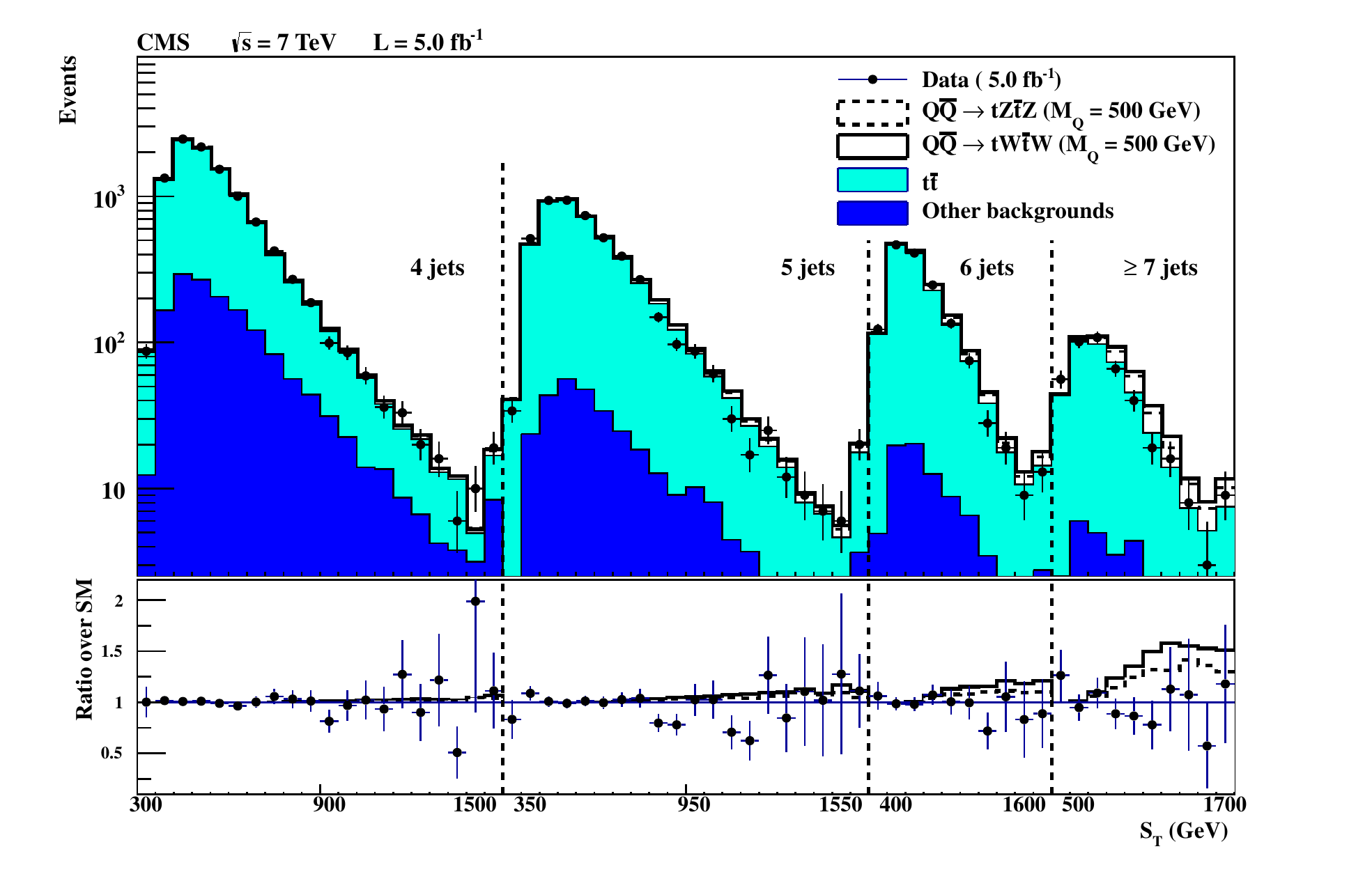}
\caption{Distribution in $S_{\mathrm T}$ for different jet multiplicities
after the maximum-likelihood fit to data. The last bin in each subfigure is the overflow
bin.The bottom plot shows the ratios of data and SM plus signal over SM. $\mathrm{Q\to tW}$
and $\mathrm{Q\to tZ}$ distributions are shown for illustrative purposes for
$M_{\mathrm Q} = 500$\GeV.}
\label{fig:fit}
\end{center}
\end{figure}

The $S_{\mathrm T}$ distributions for different jet multiplicities are combined, and
shown in Fig.~\ref{fig:fit}, after the maximum-likelihood fit to data.
No excess over the predicted SM background is observed, and we
proceed to set an upper limit on the $\mathrm{Q\overline{Q}}$ cross section.
The upper limit is extracted using a frequentist CL$_\mathrm{s}$ technique~\cite{CLs1,CLs2}
with an asymptotic approximation. The following likelihood ratio is used
as a test statistic:

\begin{equation}
t(x|\sigma) = \begin{cases}
L(x|\sigma,\hat\nu_\sigma)/L(x|\hat\sigma, \hat\nu) &\text{if } \sigma>\hat\sigma \\
1 & \text{if } \sigma\leq\hat\sigma, \end{cases}
\end{equation}

where $L(x|\sigma, \nu)$ is the likelihood that x is observed in data, given a
hypothesized value of the $\mathrm{Q\overline{Q}}$ cross section and ``nuisance''
parameters $\nu$. The values of $\sigma$ and $\nu$ for which the likelihood
reaches its maximum value are denoted $\hat\sigma$ and $\hat\nu$, respectively.
The symbol $\hat\nu_\sigma$ refers to the values of the parameters $\nu$ that
maximize the conditional likelihood for any given value of $\sigma$.

The probability to observe a value of $t$ for the likelihood ratio that is larger
than the observed value $t_{\text{obs}}$ is determined using pseudo-experiments
in which the expected numbers of signal and background events are allowed to
vary according to their statistical and systematic uncertainties. For pseudo-experiments
generated assuming a background-only hypothesis, the probability is denoted by CL$_\mathrm{b}$.
For pseudo-experiments assuming
background plus signal with a cross section $\sigma$, the probability is denoted by
CL$_{\mathrm{s+b}}(\sigma)$. The 95\% confidence level (CL) upper limit for the
$\mathrm{Q\overline{Q}}$ cross section is the value of  $\sigma$ for which
the ratio of CL$_{\mathrm{s+b}}(\sigma)$ and CL$_{\mathrm b}$, denoted as CL$_\mathrm{s}$, is 0.05.

We also verify that
the negative log likelihood is minimized at the value
corresponding to the global minimum.
We evaluate the likelihood as a function of the $\nu$ parameters
and check for secondary minima.
No such minima are observed, and the data constrain the
nuisance parameters well within their a priori assumptions.
Following the maximization of the likelihood, the dominant uncertainty is from the
matching between matrix element partons and parton showers. To quantify this effect,
the likelihood
minimization is performed excluding the parton matching systematic uncertainty.
This results in reduction of the upper limit on the signal cross section for
$M_{\mathrm Q} = 600$\GeV by 5\%, and an increase in the exclusion limit by 15\GeV.
This procedure provides an estimate of the impact of this uncertainty.

\section{Results}
\label{sec:results}

\begin{figure}[htp]
\begin{center}
\includegraphics[width=0.6\textwidth]{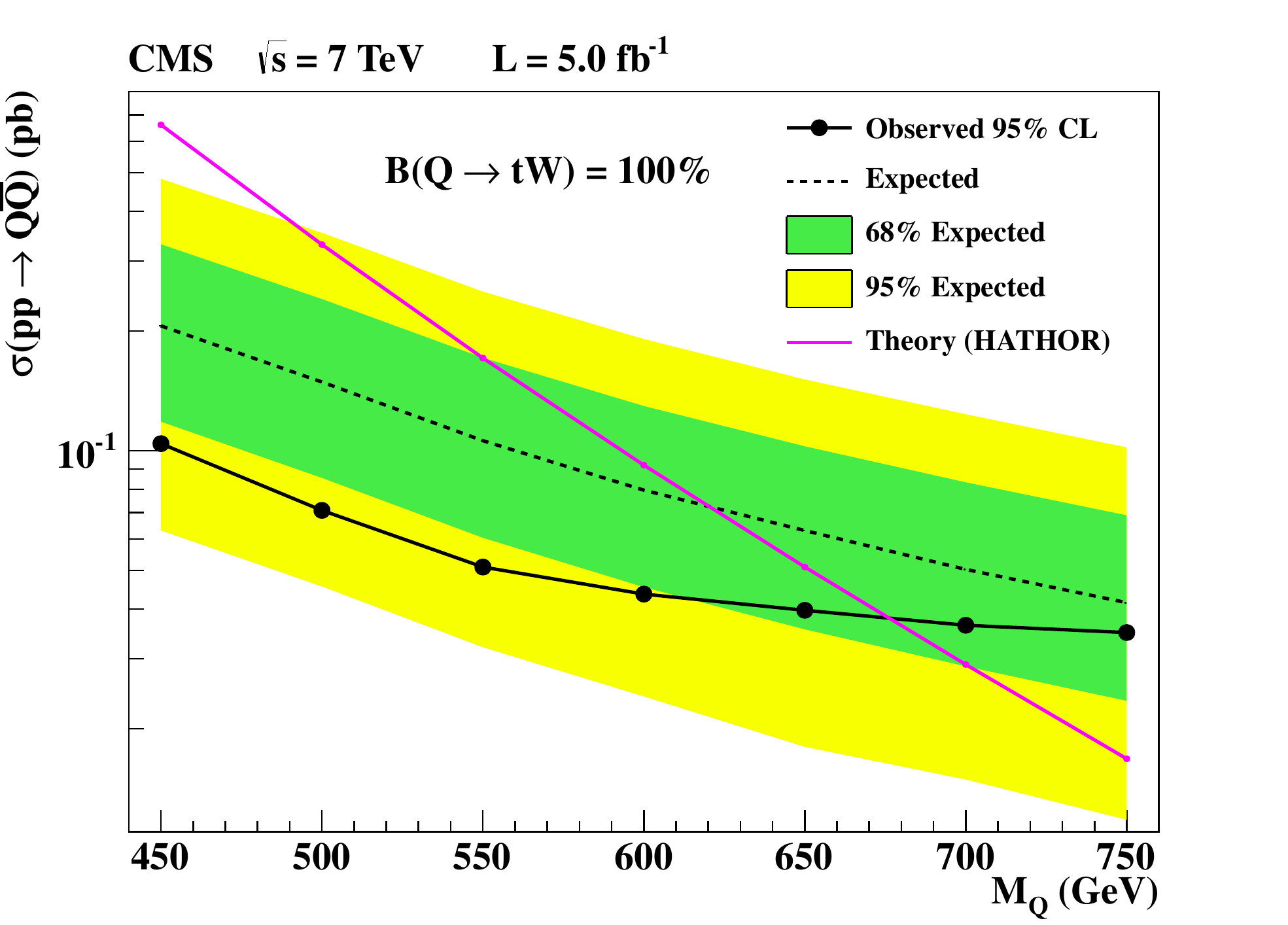}
\caption{The observed (solid line with points) and the expected
(dashed line) 95\% CL upper limits on the $\mathrm{Q\overline{Q}}$ production cross section
as a function of the heavy quark mass, $M_\mathrm{Q}$,
compared to the theoretical $\mathrm{Q\overline{Q}}$ cross section for the case of a down-type
heavy quark decaying
exclusively to $\mathrm{tW}$.}
\label{fig:limits}
\end{center}
\end{figure}

Figure~\ref{fig:limits} shows the observed and expected 95\%  CL upper
limit on the $\mathrm{Q\overline{Q}}$ production cross section, $\sigma_{\mathrm{Q\overline{Q}}}$,
for a down-type
heavy quark decaying exclusively to $\mathrm{tW}$. The lower mass limit is
determined by the value at which the observed upper limit curve for $\sigma_{\mathrm{Q\overline{Q}}}$
crosses the theoretical expectation. The observed (expected) limit corresponds to 675 (625)\GeV.

\begin{figure}[htp]
\begin{center}
\includegraphics[width=0.6\textwidth]{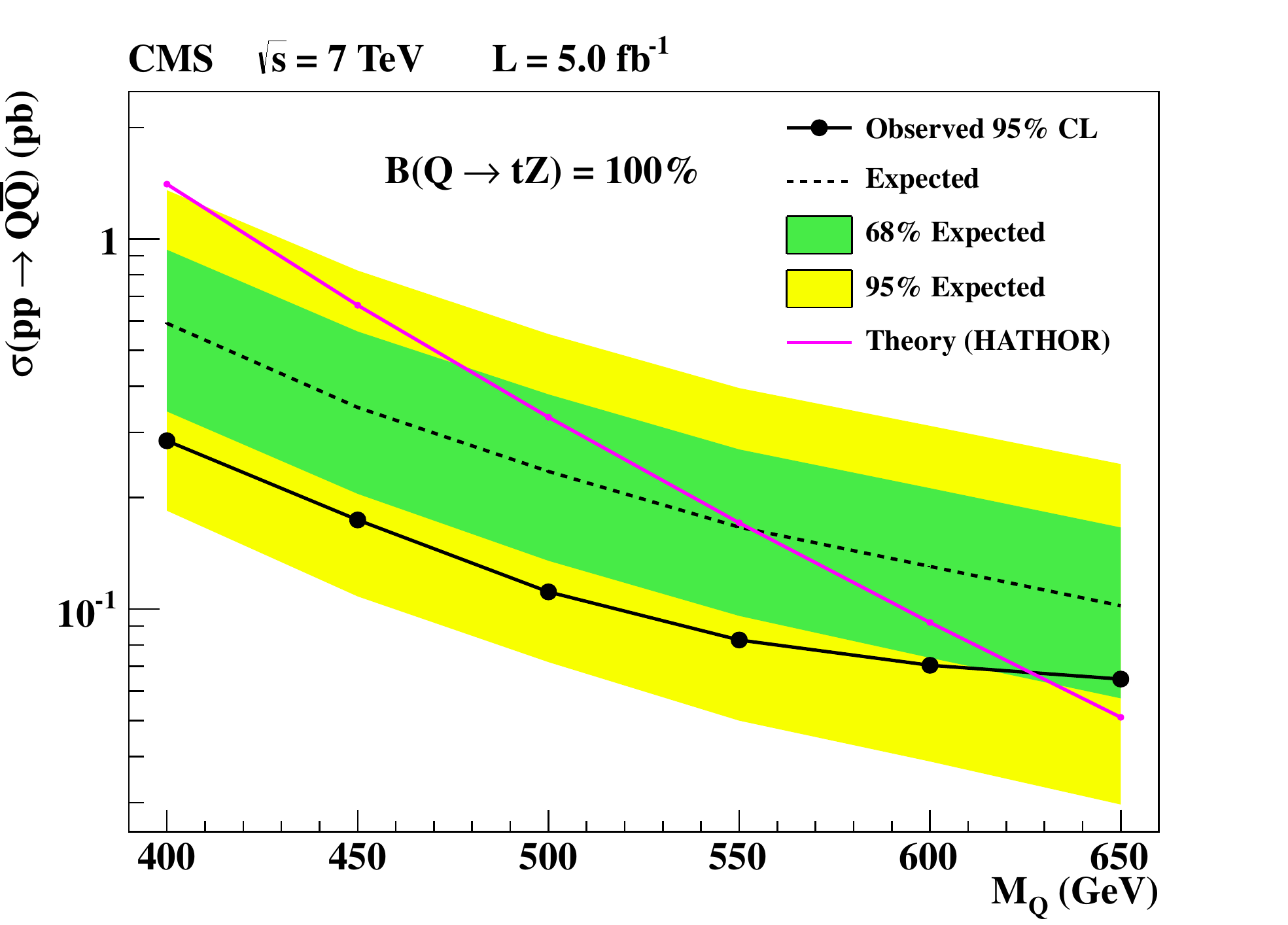}
\caption{The observed (solid line with points) and the expected
(dashed line) 95\% CL upper limits on the $\mathrm{Q\overline{Q}}$ production cross section
as a function of the heavy quark mass, $M_\mathrm{Q}$,
compared to the theoretical $\mathrm{Q\overline{Q}}$ cross section for the case of an up-type 
heavy quark decaying
exclusively to $\mathrm{tZ}$.}
\label{fig:limits_tZ}
\end{center}
\end{figure}

Figure~\ref{fig:limits_tZ} shows the observed and expected 95\% CL upper limit on the
$\sigma_{\mathrm{Q\overline{Q}}}$ as a function of quark mass ($\mathrm{Q}$)
for an up-type heavy quark decaying exclusively to $\mathrm{tZ}$. The observed
(expected) limit corresponds to 625 (550)\GeV.

Several cross checks have been performed to investigate the difference between the observed
and expected limits. We studied several models of the $\ttbar$ $S_{\mathrm T}$ spectrum by using
different generators, such as {\PYTHIA} and {\POWHEG}.  All of the generators provide
results similar to {\MADGRAPH} within their systematic uncertainties. We also
studied different models of the $\ttbar$ $S_{\mathrm T}$ spectrum by changing internal parameters
in {\MADGRAPH}, such as the renormalization and factorization scales and the parameters
responsible for matching jets originating from matrix element partons to their showers.
We determine that a change of
the matching parameters by one standard deviation from their nominal values provides
good agreement between the simulated and observed spectrum of the $S_{\mathrm T}$ distribution, which
can be accommodated in the fit because of the relatively weak dependence of the minimum on
this parameter. The dependence of the $S_{\mathrm T}$ distribution for $\ttbar$ background is covered
by the systematic uncertainties included in the fit of the model to data.

\section{Summary}
\label{sec:summary}

A search for pair-produced new heavy quarks $\mathrm{Q\overline{Q}}$ decaying
exclusively to $\mathrm{tW\overline{t}W}$ or to $\mathrm{tZ\overline{t}Z}$
is performed in lepton $+$ jets events. The analysis is based on a data sample
of proton-proton collisions at $\sqrt{s} = 7$\TeV corresponding to an
integrated luminosity of 5.0\fbinv. Events are selected
requiring an electron or a muon, missing transverse momentum, and at least four jets, one of which
is identified as a bottom jet. A combined fit is performed to the scalar sum of the transverse momenta of
all final reconstructed objects as a function of jet multiplicity. No significant deviations from
SM expectations are found, and upper limits on the production cross section of $\mathrm{Q\overline{Q}}$
as a function of a heavy quark mass are computed. Assuming a strong production mechanism for both
signal models, down-type quarks decaying exclusively to $\mathrm{tW}$ with
masses below 675\GeV and up-type 
quarks decaying exclusively to $\mathrm{tZ}$
with masses below 625\GeV are excluded at 95\% CL. These are the most stringent limits
to date.

\section*{Acknowledgements}
We congratulate our colleagues in the CERN accelerator departments for the excellent performance of the LHC machine. We thank the technical and administrative staff at CERN and other CMS institutes, and acknowledge support from: FMSR (Austria); FNRS and FWO (Belgium); CNPq, CAPES, FAPERJ, and FAPESP (Brazil); MES (Bulgaria); CERN; CAS, MoST, and NSFC (China); COLCIENCIAS (Colombia); MSES (Croatia); RPF (Cyprus); MoER, SF0690030s09 and ERDF (Estonia); Academy of Finland, MEC, and HIP (Finland); CEA and CNRS/IN2P3 (France); BMBF, DFG, and HGF (Germany); GSRT (Greece); OTKA and NKTH (Hungary); DAE and DST (India); IPM (Iran); SFI (Ireland); INFN (Italy); NRF and WCU (Korea); LAS (Lithuania); CINVESTAV, CONACYT, SEP, and UASLP-FAI (Mexico); MSI (New Zealand); PAEC (Pakistan); MSHE and NSC (Poland); FCT (Portugal); JINR (Armenia, Belarus, Georgia, Ukraine, Uzbekistan); MON, RosAtom, RAS and RFBR (Russia); MSTD (Serbia); MICINN and CPAN (Spain); Swiss Funding Agencies (Switzerland); NSC (Taipei); TUBITAK and TAEK (Turkey); STFC (United Kingdom); DOE and NSF (USA).

\bibliography{auto_generated}   
\cleardoublepage \appendix\section{The CMS Collaboration \label{app:collab}}\begin{sloppypar}\hyphenpenalty=5000\widowpenalty=500\clubpenalty=5000\textbf{Yerevan Physics Institute,  Yerevan,  Armenia}\\*[0pt]
S.~Chatrchyan, V.~Khachatryan, A.M.~Sirunyan, A.~Tumasyan
\vskip\cmsinstskip
\textbf{Institut f\"{u}r Hochenergiephysik der OeAW,  Wien,  Austria}\\*[0pt]
W.~Adam, E.~Aguilo, T.~Bergauer, M.~Dragicevic, J.~Er\"{o}, C.~Fabjan\cmsAuthorMark{1}, M.~Friedl, R.~Fr\"{u}hwirth\cmsAuthorMark{1}, V.M.~Ghete, J.~Hammer, N.~H\"{o}rmann, J.~Hrubec, M.~Jeitler\cmsAuthorMark{1}, W.~Kiesenhofer, V.~Kn\"{u}nz, M.~Krammer\cmsAuthorMark{1}, I.~Kr\"{a}tschmer, D.~Liko, I.~Mikulec, M.~Pernicka$^{\textrm{\dag}}$, B.~Rahbaran, C.~Rohringer, H.~Rohringer, R.~Sch\"{o}fbeck, J.~Strauss, A.~Taurok, W.~Waltenberger, G.~Walzel, E.~Widl, C.-E.~Wulz\cmsAuthorMark{1}
% ...
\vskip\cmsinstskip
\dag:~Deceased\\
1:~~Also at Vienna University of Technology, Vienna, Austria\\
% ...

\end{sloppypar}
\end{document}